\documentclass[]{raa}            

\usepackage{graphicx,times}             
\usepackage{graphics}
\usepackage{epsfig}
\usepackage{multirow}
\usepackage{natbib}
\usepackage{amssymb,amsmath}
\bibpunct{(}{)}{;}{a}{}{,}

\usepackage[a4paper=true,dvipdfm=true,pagebackref=true]{hyperref}
\hypersetup{colorlinks = true, linkcolor = green, anchorcolor = red, citecolor = blue, filecolor = red, pagecolor = red, urlcolor = red}

\begin{document}

   \title{Investigation of near-contact semi-detached binary \\
W UMi through observations and evolutionary models
}

   \volnopage{Vol.0 (20xx) No.0, 000--000}      
   \setcounter{page}{1}          

   \author{F. Soydugan
      \inst{1,2}
   \and E. Soydugan
      \inst{1,2}
   \and F. Ali\c{c}avu\c{s}
      \inst{1,2}
   }

   \institute{\c{C}anakkale Onsekiz Mart University, Faculty of Arts and Sciences, Department of Physics, 17020, \c{C}anakkale, Turkey; {\it fsoydugan@comu.edu.tr} \\
\and
    \c{C}anakkale Onsekiz Mart University, Astrophysics Research Center and Ulup{\i}nar Observatory, 17020, \c{C}anakkale, Turkey\\
\vs\no
   {\small Received~~20xx month day; accepted~~20xx~~month day}}

\abstract{ W UMi is a near contact, semi-detached, double-lined
eclipsing binary star with an orbital period of 1.7 days.
Simultaneous analysis of new \emph{BVR} multi-color light curves and
radial velocity data yields the main astrophysical parameters of the
binary and its component stars. We determined mass and radius to be
\emph{M$_{1}$}=3.22$\pm$0.08 \emph{M$_{\odot}$},
\emph{R$_{1}$}=3.63$\pm$0.04 \emph{R$_{\odot}$} for the primary star
and \emph{M$_{2}$}=1.44$\pm$0.05 \emph{M$_{\odot}$},
\emph{R$_{2}$}=3.09$\pm$0.03 \emph{R$_{\odot}$} for the secondary
star. Based on analysis of mid-eclipse times, variation in the
orbital period is represented by a cyclic term and a downward
parabola. Mass loss from the system is suggested for a secular
decrease (-0.02 s yr$^{-1}$) in the period. Both the mechanisms of a
hypothetical tertiary star orbiting around W UMi and the surface
magnetic activity of the less massive cooler companion were used to
interpret periodic changes. Observational parameters were found to
be consistent with binary stellar evolution models produced in the
non-conservative approach of \texttt{MESA} at a higher metallicity
than the Sun and an age of about 400 Myr for the system. Evidence
that the system is rich in metal was obtained from spectral and
kinematic analysis as well as evolution models. W UMi, a high mass
ratio system compared to classical semi-detached binaries, is an
important example since it is estimated from binary evolutionary
models that the system may reach its contact phase in a short time
interval.
\keywords{binaries: eclipsing, stars: fundamental parameters,
stars:evolution, stars: individual: W UMi} }

   \authorrunning{F. Soydugan, E. Soydugan \& F. Ali\c{c}avu\c{s}}            
   \titlerunning{A near-contact semi-detached binary W UMi }  

   \maketitle
%
%

\section{Introduction}           
\label{sect:intro}

In the field of astrophysics, the role and importance of
double-lined eclipsing binaries (EBs) is well known. EBs provide
accurate determination of absolute parameters, mainly the radius and
mass of the components from simultaneous analysis of light and
radial velocity curve data. Moreover, they are good indicators of
distance.

Semi-detached Algol binaries (SDBs) consist of a less massive,
sub-giant or giant secondary component of F-G-K spectral type and a
hotter, more massive, main-sequence  primary component which is B-A
spectral-type. In this type of EBs, the secondary companion, which
fills its Roche lobe, can display the loss of angular momentum and
mass transfer to the more massive companion. In long period Algols
with orbital periods (\emph{P}) longer than 5-6 days, we can expect
permanent accretion disc because of the accretion structure of the
primary component \citep*{b14}. On the other hand, Algols with
\emph{P}$<$3-4 days show temporary weak accretion structures or the
evidence of a mass transfer process cannot be observed. SDBs are
important because they display meaningful astrophysical processes
and are useful for studying binary star evolution when taking into
account the effects of mass transfer and loss.

Near-contact binaries (NCBs) are within the intermediate class
according to Roche classification \citep{b38}. Semi-detached Algols
may be also a member of SD2-type of NCBs, if the massive companions
are very close to their inner Roche lobes \citep{b57}. NCBs can be
investigated to study the evolutionary process of a possible
transition from detached binary to semi-detached binary, and also
from semi-detached binary to contact binary. In this study,
semi-detached binary W UMi is a SD2 type NCB candidate system.

W UMi, (HIP 79069, BD+86 244, SAO 2692) is a classical Algol-type
binary system with an orbital period of about 1.701 day, was
discovered by \citet{b2}. The first spectroscopic study of W UMi was
carried out by \citet{b11} who measured the radial velocity (RV)
values associated with only the primary component and estimated the
spectral type to be A4 for the component. In a later spectroscopic
study by \citet{b37}, because of determining lines belonging only to
the hotter primary component, W UMi was again classified as single
line spectroscopic binary. Recently, \citet{b24} presented a
spectroscopic and photometric study, which includes RVs of the
primary and secondary stars and an orbital solution of the system.

\emph{UBV} light curves (LCs) of W UMi were acquired by \citet{b4}
and solved by Russell-Merrill method. The same photometric data were
re-analyzed by \citet{b22}. Recent analysis of the LCs for the
system was reported by \citet{b24}, in which the mass ratio
(\emph{q}) of the components was determined and used to be q=0.399.
In addition, the temperature of the massive star was found to be
9300$\pm$90 K using spectral analysis \citep{b24}. Therefore, the
basic astrophysical parameters in that study were estimated as
different from previous studies \citep[e.g.][]{b22,b5}. \citet{b25}
firstly reported changes in the orbital period of the target.
Orbital period variations of W UMi were interpreted as being in
continuous decrease and also experiencing sudden changes as a result
of mass loss by \citet{b23}. Lastly, \citet{b16} announced a
decrease and also a cyclic variation in the period. The cyclic
changes were interpreted with a probable third body in the system
specified as a M dwarf star.

The present study presents an analysis of new photometric and
spectroscopic data and also the evolutionary status of W UMi, taking
into account binary effects. Section 2 includes information on
photometric and spectroscopic observations. RV values of the
companions, properties of the system's orbit, and also the model
atmosphere application are given in Section 3. Simultaneous analysis
of the photometric multi-color LCs and RV data of the components are
reported in Section 4. In the next section, we focus on probable
factors causing variations in the period of the target. The
following section includes the main astrophysical properties of the
components and also evolutionary scenarios for W UMi. Finally, a
discussion about the results and conclusions is presented.


\section{Photometric and Spectroscopic Observations}
\label{sect:Obs}

The differential \emph{BVR} LCs of W UMi were observed during April
- May 2012 on 9 nights at the Observatory of \c{C}anakkale Onsekiz
Mart University using the 30-cm diameter Schmidt-Cassegrain
telescope with STL-1001E camera. We chose GSC 4655-00320 (V =
9$^{m}$.79) and GSC 4651-00275 (V = 11$^{m}$.03) as the comparison
and check stars and no light changes were indicated for these stars
during observations. We reduced the CCD frames obtained from each
night's observation using C-Munipack code
(http://integral.sci.muni.cz/). The mean errors were calculated as
about 0$^{m}$.01 for all filters. The initial epoch T0 (HJD) =
2452500.3960 and orbital period P = 1.7011371 days taken from
\citet{b17} were used to calculate the orbital phases in the LCs.

Spectroscopic data of the target were acquired with a 0.91 meter
diameter telescope equipped with e\'{c}helle spectrograph (FRESCO)
at Catania Astrophysics Observatory. FRESCO has a resolving power of
about about 20 000 {\AA}. A CCD camera of 1024$\times$1024 pixels
(size 24$\times$24 $\mu$m) with a thinned back illuminated SITe was
utilized. The spectral data were recorded in 19 orders from about
4300 to 6650 $\AA$. Signal to noise (S/N) ratio showed variations
between 30-50 in the H$\alpha$ continuum region. In the
spectroscopic observations of W UMi, Vega (A0V) and $\alpha$ Boo
(K1.5 III) were selected as the standard stars for the hotter
primary and cooler secondary companions, respectively. According to
atmospheric conditions, exposure times between 2500 and 3300 seconds
were given for a total of 16 spectra obtained. For the data
reduction, we used the ECHELLE task of IRAF code, including the
following steps: light subtraction from background, division by a
flat field obtained by a halogen lamp, wavelength calibration with
the spectrum of a Th-Ar lamp, and normalization of the spectra to
the continuum.

\section{Spectroscopic Analysis}

\subsection{RV Measurements and Orbit
Properties}

In the spectroscopic studies by \citet{b37} and \citet{b11}, merely
the RV measurements of the primary component were determined. W UMi
was known as a single-lined binary until the study of \citet{b24},
who measured the RV values of both companions and reported orbital
parameters. In order to test the orbital parameters found by
\citet{b24}, we aimed to calculate the orbital solution using our
own data by measuring RVs of both components with many more
absorption lines.

The radial velocity measurements of W UMi were made using the IRAF
task FXCOR which includes the cross-correlation technique, commonly
used for this aim. RVs of the hotter, primary component and cooler,
secondary component were calculated from 12 and 11 spectra,
respectively. We eliminated 2 spectra from the 16 spectra due to
large errors while the RVs were designated for the primary
component. The standard stars Vega (A0V, RV = -13.9 km s$^{-1}$) for
the primary star and $\alpha$ Boo (K1.5III, RV = -5.9 kms$^{-1}$)
for the secondary star were selected as templates for application of
the cross-correlation method. Na I D2 and H$\alpha$ lines were
excluded in the spectroscopic analysis in order not to be affected
by mass transfer and magnetic activity. Furthermore, we did not use
the regions affected by telluric lines. Finally, the wavelength
region between 4400 and 6000 $\AA$ was preferred to measure RV
values of the hotter and cooler companions.

The RVs of the components and their errors determined from
spectroscopic analysis are listed in Table 1. The weighted averages
of the RVs were determined from the cross-correlation of each order
of W UMi spectra used, with the corresponding order of the spectra
of standard stars. Calculation methods for the weighted means of the
RVs and their errors can be found in several studies
\citep[e.g.][]{b8,b42}.

The secondary component is dimmer than the primary component in the
W UMi system. Therefore, the error values of the RVs of the
secondary component were found to be larger. The error values
computed from RV measurements are approximately 5 km s$^{-1}$ and 10
km s$^{-1}$ for the hotter and cooler components, respectively. The
accordance of observational points with theoretical curves is
presented in Figure 1. The spectroscopic orbital parameters given in
Table 2 were used as input parameters in the simultaneous analysis
of light and RV curves data. According to the orbital parameters, it
is clear that the semi-amplitude for the RV changes in the secondary
component (\emph{K}$_{2}\approx$197$\pm$3 km $s^{-1}$) and mass
ratio of the companion stars (\emph{q}=0.460) were found to be
different from those given by \citet{b24}.

\begin{table}
 \begin{center}
  \caption{Radial velocities and their errors for hot and cool components of W UMi.}
  \begin{tabular}{@{}cccr@{}}
  \hline\hline
HJD         & Orbital  & V$_{1}$           & V$_{2}$~~~~~       \\
24 50000+   & Phase    & (km s$^{-1}$)     & (km s$^{-1}$)  \\
  \hline
3160.4684  & 0.0183    & --~~~~~~          & $7.6 \pm$ 11.5 \\
3126.5094  & 0.0558    & --~~~~~~          &$58.4 \pm$ 12.8    \\
3155.4744  & 0.0826    & $-58.3 \pm$ 4.3   &$90.9 \pm$ 9.7    \\
3167.5015  & 0.1527    & $-89.7 \pm$ 5.6   & --~~~~~~ \\
3184.5565  & 0.1783    & $-92.5 \pm$ 4.7   &$147.3 \pm$ 9.4    \\
3157.5005  & 0.2737    & $-113.8 \pm$ 4.3  & $182.4 \pm$ 13.2 \\
3203.5845  & 0.3638    & $-78.9 \pm$ 5.6   &--~~~~~~   \\
3125.5839  & 0.5118    & $-6.6 \pm$ 2.5    &--~~~~~~   \\
3127.5672  & 0.6776    & $56.5 \pm$ 4.5    & $-189.1 \pm$ 12.6  \\
3185.4074  & 0.6785    & $64.8 \pm$ 5.2    &$-192.6 \pm$ 12.3    \\
3202.5581  & 0.7604    & $73.4 \pm$ 6.1    &$-213.5 \pm$ 11.1    \\
3151.5535  & 0.7778    & $79.3 \pm$ 6.8    &$-201.3 \pm$ 10.6    \\
3170.4324  & 0.8756    & $54.1 \pm$ 6.6    &$-160.5 \pm$ 8.3    \\
3124.5607  & 0.9103    & $39.8 \pm$ 6.1    &$-131.0 \pm$ 9.2    \\
\hline
\end{tabular}
\end{center}
\end{table}

\begin{table}
  \caption{Orbital parameters for W UMi.}
  \label{tab:orbit}
  \begin{center}
           \begin{tabular}[h]{lll}
      \hline\hline \\[-8pt]
          Parameter    & &    Value                     \\
      \cline{1-1}\cline{3-3}\\
      \emph{T$_{0}$} (HJD)          & & 2452500.3960\footnotemark[1]  \\
      \emph{P$_{orb}$ }(day)        & & 1.7011371\footnotemark[1]     \\
      \emph{V$_{\gamma}$ }(km s$^{-1}$) & & -15.3 $\pm$ 1.1                \\
      \emph{K$_{1}$ } (km s$^{-1}$) & & 90.4 $\pm$ 1.5               \\
      \emph{K$_{2}$ } (km s$^{-1}$) & & 196.6 $\pm$ 2.6              \\
    \emph{a$_{1}$ sin i} (10$^{6}$\,km) &  & 2.11 $\pm$ 0.05         \\
    \emph{a$_{2}$ sin i} (10$^{6}$\,km) &  & 4.60 $\pm$ 0.10         \\
      \emph{M$_{1}$ sin$^{3}$\,i} (\emph{M$_{\odot}$}) & &  2.85 $\pm$ 0.16 \\
      \emph{M$_{2}$ sin$^{3}$\,i} (\emph{M$_{\odot}$}) & &  1.31 $\pm$ 0.06 \\
       \hline
     \end{tabular}\\
      \footnotemark[1]{\citet{b17}}\\
      \end{center}
      \end{table}

\begin{figure}
\begin{center}
\includegraphics*[width=8.3cm,height=5.8cm,scale=1.0,angle=000]{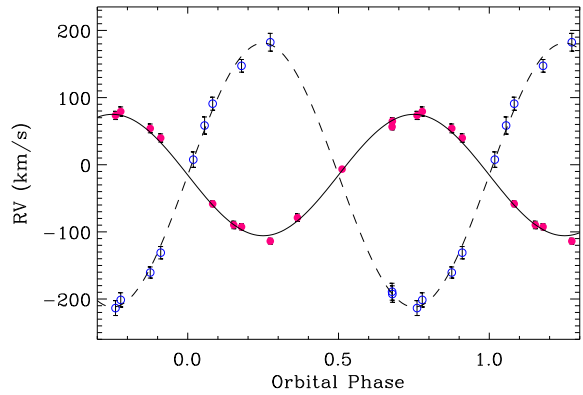}
\caption{RV variations of components of W UMi versus orbital phase.
RV values of the primary and secondary stars are shown by filled and
open circles, respectively.} \label{fig1}
\end{center}
\end{figure}

\subsection{Model Atmosphere Application}

In order to model the spectra of W UMi, especially the dominant more
massive, hotter companion, the observed spectrum at the orbital
phase of about 0.5 was used. We preferred the spectrum given by
\citet{b24} since the spectrum is around the secondary minimum and
also has higher resolution. It is almost impossible to determine the
atmospheric properties of the cooler secondary companion of W UMi
because of its low light contribution (5-10$\%$) to the total light
of the target in the optical region.

Our main goal in the spectral synthesis was primarily to test
whether the metallicity of the primary component is different from
that of the Sun since binary evolution models indicate that the
system is rich in metal as seen in Section 6.

For modeling of the spectrum, the code of Spectroscopy Made Easy
(\texttt{SME}) was chosen \citep{b49}. The software can be used for
fitting the observed spectrum with the spectral synthesis method.
Atmospheric parameters of the hotter component of W UMi were
determined by interpolation of the ATLAS model atmospheres
\citep{b18} included in \texttt{SME} software. We have used the line
information from the Vienna Atomic Line Database
\citep[VALD;][]{b30}. The spectral region 4460-4560 $\AA$, including
several absorption metallic lines and also H$\alpha$ line were used
for the application. The temperature (\emph{T$_{eff,1}$}=9310$\pm$90
K) and projected rotational velocity (\emph{v$_{1}$ sin i}=107$\pm$5
km s$^{-1}$) of the hotter companion were derived. During the model
atmosphere analysis, the value of surface gravity of the primary
component (log \emph{g$_{1}$}=3.83$\pm$0.02), which was calculated
from the results obtained from the simultaneous analysis of the
light and radial velocity curves, was taken as constant in order to
avoid parameter degeneration. All atmosphere parameters were found
in accordance with the values given by \citet{b24}. On the other
hand, although \citet{b24} made spectral analysis under the
assumption solar metallicity, our results displayed higher
metallicity than the Sun for the system.

Iron content was used to estimate metallicity of the primary
component since the [Fe/H] ratio is easy to measure in the optical
spectra. We measured the [Fe/H] ratio to be about +0.5$\pm$0.2 from
the spectra obtained at the orbital phase of about 0.5 (HJD
245284.3062) by \citet{b24}. The measured [Fe/H] ratio was converted
to metallicity (Z) value as Z$\approx$0.05. Comparison between the
observed spectrum of W UMi and the synthetic spectrum for solar
metallicity and Z=0.05 can be seen in Figure 2. This Z value
indicates that the primary companion may be a metal-rich star
compared to the Sun, which coincides with the evolution models in
this study.

\begin{figure}[!ht]
\begin{center}
\includegraphics[width=8.3cm,height=10.5cm,scale=1.0,angle=000]{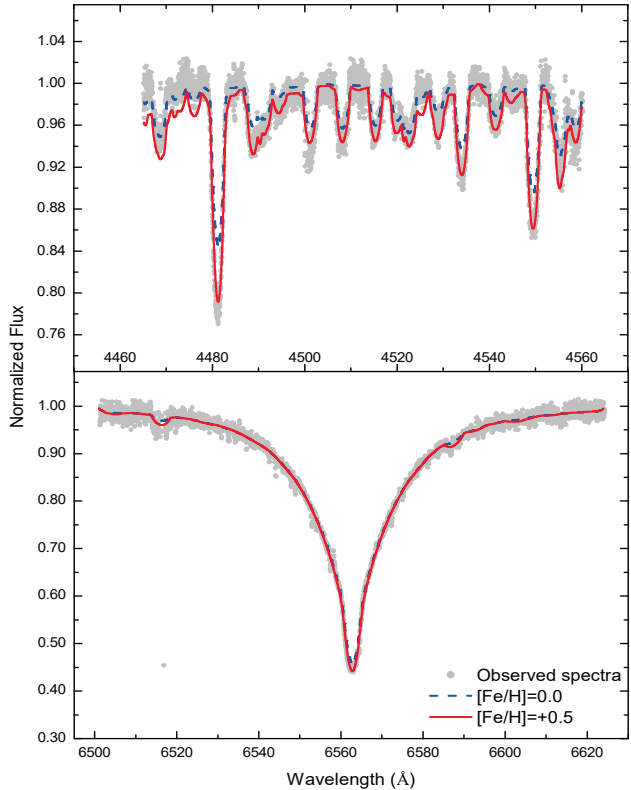}
\caption{Synthetic spectrum calculated by \texttt{SME} for
[Fe/H]=0.0 (blue dashed line) and [Fe/H]=+0.5 (red dashed-dotted
line) together with observed spectrum (gray dots) in wavelength
region 4460-4560 $\AA$ (upper panel) and around region of H$\alpha$
line (bottom panel).} \label{fig1}
\end{center}
\end{figure}

\section{Simultaneous solution of \emph{BVR} light and RV curves}
\label{sect:data}

The RV values of the companions and multi-color LCs of W UMi
acquired in this study were simultaneously analyzed by
Wilson-Devinney (WD) code \citep{b48}. All magnitude values obtained
from observations in \emph{B,V} and \emph{R} bands were normalized
to unity around the orbital phase of 0.25. In the WD solution, we
used 2930, 2905 and 2890 observational points in \emph{B,V} and
\emph{R} filters, respectively. All observational points were of the
same weight during the analysis. In order to decide accurate MOD in
the solution, firstly, we tried the MOD2 given for detached systems.
As the surface potential of the less massive, secondary star reached
its Roche limit, the analysis was continued and the process was
completed with MOD5 corresponding to semi-detached systems.

During the solution, we fixed some parameters as constant while
others were used as free. Fixed parameters were the limb-darkening
coefficients from \citet{b50}, the values of bolometric albedos as
1.0 and 0.5 from \citet{b34}, the gravity darkening coefficients for
radiative atmospheres to be 1.0 from \citet{b47} and for convective
atmospheres they were 0.32 from \citet{b20}. The rotational
parameters (\textit{F}$_{1}$) and (\textit{F}$_{2}$) of both two
components were selected as 1.0, assuming synchronous rotation for
both companions. A circular orbit (\textit{e}=0) was accepted as
obtained from the solution of the spectroscopic orbit. The free
parameters were the inclination (\textit{i}) of the system's orbit,
temperature of the cooler, secondary (\textit{T}$_{2}$), surface
potential of the hotter, primary ($\Omega_{1}$), phase shift, mass
ratio (\textit{q}), semi-major axis (\textit{a}) of the orbit,
radial velocity (\textit{V}$_{\gamma}$) of the system's mass center
and the fractional value of the primary luminosity
(\textit{L}$_{1}$) during iterations.

In the analysis, the leading parameters were the temperature of the
hotter star and mass ratio of the companion stars. The temperature
of the more massive, primary companion was chosen as being 9310 K as
given by \citet{b24}, which is tested in this study by applying
spectral synthesis around the H$\alpha$ line region and also in the
wavelength range 4460-4560 $\AA$. Because of the difference between
the spectroscopic mass ratio obtained in this study and the q value
determined by \citet{b24}, photometric q-search was performed using
the light curves obtained in this study. As a result of this
scanning, the minimum value of $\sum(\emph{O-C})^{2}$ (the weighted
sum of the squared residuals) was obtained around q=0.44, which is
close to the spectroscopic q value in this study. Therefore, the
mass ratio value (\textit{q}) of 0.460 listed in Table 2 and
obtained from the orbital solution was used as the input parameter
in the solution. In the results of the photometric modeling, the
mass ratio obtained was 0.447, which is very close to the
spectroscopic mass ratio value.

In order to test the third light contributing to the total light of
W UMi, a third light parameter (\emph{l$_{3}$}) was chosen as a free
parameter. The third light was accepted to be zero during the
iterations since a meaningful value was not obtained for
\textit{l}$_{3}$. The physical and geometrical parameters derived
from simultaneous analysis of the RVs and multi-color LCs are listed
in Table 3. Comparison of observational points and model curves are
shown in Figure 3. Roche lobes together with the binary display in
accordance with parameters determined from the analysis of
\emph{BVR} LCs indicate that while the secondary component fills the
Roche lobe, the primary companion fills about 85 \% of its inner
lobe (see Fig. 7). Although W UMi is classified as a classic
Algol-type eclipsing binary, it is a candidate showing
characteristics of a near contact semi-detached binary system,
according to results in Table 3 and also Roche geometry in Fig. 7.

\begin{table}
\caption{Parameters derived from analysis of \emph{BVR} multi-color
LCs and RV data for W UMi.} \label{table3}
    $$
        \begin{tabular}{lc}
    \hline
Parameter  & Value                                                             \\
    \hline

\textit{a} (\emph{R$_{\odot}$}) & 10.01(6) \\
\textit{V}$_{\gamma}$ (km s$^{-1}$) & -15.05 (1.2) \\
\textit{i} ($^{\circ}$)              & 83.57(18)                            \\
\textit{T}$_{1}$ (K)                 & 9310$^{a}$                              \\
\textit{T}$_{2}$ (K)                 & 5240(30)                             \\
$\Omega_{1}$                        & 3.266(16)                             \\
$\Omega_{2}$                        & 2.773                               \\
Phase shift                         & 0.0001(1)                            \\
\textit{q}                          & 0.447(5)                             \\

\textit{L}$_{1}$/(\textit{L}$_{1}$+\textit{L}$_{2}$) & 0.947(8) (\emph{B})  \\
                                                     & 0.920(5) (\emph{V})  \\
                                                     & 0.895(3) (\emph{R})  \\
\textit{L}$_{2}$/(\textit{L}$_{1}$+\textit{L}$_{2}$) & 0.053 (\emph{B})      \\
                                                     & 0.080 (\emph{V})      \\
                                                     & 0.105 (\emph{R})      \\
{\textit{r}}$_{1}$ (mean)                             & 0.363(1)     \\
{\textit{r}}$_{2}$ (mean)                             & 0.309(1)     \\

\noalign{\smallskip} \hline \noalign{\smallskip}
\end{tabular}
    $$
\begin{center}
    \begin{minipage}{60mm}
$^{a}$Adopted from \citet{b24}
\end{minipage}
\end{center}
\end{table}

\begin{figure}
\begin{center}
\includegraphics*[width=8.3cm,height=6.5cm,scale=1.0,angle=000]{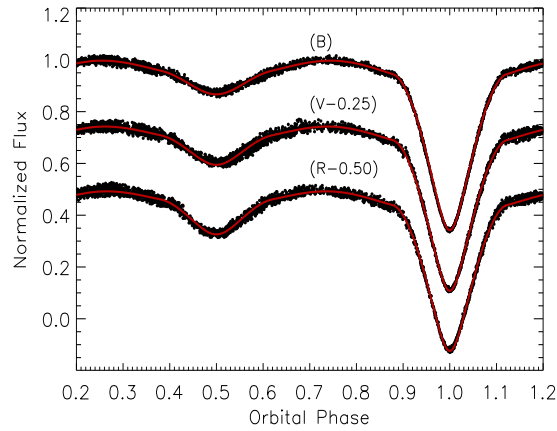}
\caption{\emph{BVR} light curves of W UMi along with theoretical
fits.} \label{fig3}
\end{center}
\end{figure}

\section{Orbital period variations}

Classical Algol-type eclipsing binary stars or SDBs may indicate
changes in their orbital periods with different forms. This can be
seen for many systems in the O-C catalogue of eclipsing binary
systems \citep{b15}. Some SDBs show upward or downward parabolic
changes together with cyclic variations or not, although an increase
in the orbital period can be expected due to their evolutionary
status. This reveals that different physical processes (e.g. mass
loss and/or transfer, magnetic activity, existing unseen components)
can be effective at the same time or some can be more dominant from
time to time.

In this study, the orbital period behavior of semi-detached binary W
UMi was studied over a time interval of about a century. In addition
to three minima time obtained in this study which are given in Table
4, 201 eclipse times (172 visual (v), 2 photographic plate (p) and
27 photoelectric (pe) and CCD) were used in the analysis. The data
were taken from the "O-C Gateway" (http://var2.astro.cz/ocgate/) and
the literature. In this study, analysis was carried out based on O-C
data as used for many eclipsing binaries
\citep[e.g.][]{b43,b53,b52}.

As noted by \citet{b16}, parabolic and cyclic changes in the O-C
graph of the target system are clearly seen. Therefore, the O-C data
represents a cyclic term superimposed on a parabola. A secular
decline in the orbital period of semi-detached binaries does not
coincide with the evolutionary status of these systems since their
less massive components are filling their Roche lobes and
transferring their mass to the hotter, more massive companion.
However, the dominant mechanism may be mass and angular momentum
loss from the binary. The cyclic variation can be the result of a
tertiary star in the close pair or the magnetic cycle of the
secondary component with convective outer layers. These two possible
mechanisms were evaluated in this study to interpret cyclic changes
in the O-C diagram.

\begin{table}
\caption{Obtained minima times in this study for W UMi.}
\label{table3}
    $$
        \begin{tabular}{cccc}
    \hline
JD. Hel.+2400000  & Error    & Min.  & Filters    \\
    \hline
56046.4245 & 0.0008 & II & BVR  \\
56051.5146 & 0.0006 & II & BVR   \\
56052.3721 & 0.0002 & I  & BVR   \\
\noalign{\smallskip} \hline \noalign{\smallskip}
\end{tabular}
    $$

\end{table}

Firstly, O-C data of W UMi is represented by a downward parabola
plus light-time effect (LITE) using the following equation:

\begin{equation}
T = T_{0} + E \times P+ Q \times E^{2} + \Delta t
\end{equation}
where \emph{T$_{o}$} (reference time for primary eclipse) and
\emph{P} (orbital period) are the light elements of W UMi, while the
time delay resulting from an unseen star in the close binary is
represented by ${\Delta}$\emph{t}. The LITE equation as described as
a time delay, which is given by \citet{b12} is commonly used for the
analysis of changes in orbital period \citep[e.g.][]{b40,b41,b55}.

The parameters derived from the analysis applying the quadratic term
and LITE to O-C data are reported in Table 5. Distribution of O-C
values, applied model fits and also residuals from the theoretical
representation are indicated versus years and also epoch numbers in
Figure 4. LITE parameters in Table 5 indicate that W UMi has an
eccentric orbit around the mass-center of the triple system with a
period of $\approx$71 yrs. The projected distance of the close
binary's mass center to the mass-center of triple system
($a_{12}sini'$) was calculated as 2.99 AU. These values of the two
parameters allow us to estimate the mass function of the possible
tertiary star to be $f(M_{3})\approx0.0053 M_\odot$. Therefore, the
minimum mass of the unseen companion was calculated to be 0.49
$M_\odot$. The amplitude of RV changes of the mass-center of W UMi,
relative to that of the third-body system, would be 2.6\,\,km
s$^{-1}$, which can be detectable spectroscopically if spectra can
be obtained with a high resolution and high signal-to-noise ratio.
We found the maximum angular separation between the tertiary
component and the close binary to be approximately 0$''$.12 for a
distance of 414 pc which was taken from the Gaia-DR2 database
\citep{b9,b10}.

\begin{table}[!ht]
\caption{Parameters and their errors obtained from analysis of O-C
data for W UMi.} \label{table4}
    $$
        \begin{tabular}{ll}
\noalign{\smallskip} \hline \noalign{\smallskip}
Parameters                      & Value              \\
\noalign{\smallskip} \hline \noalign{\smallskip}
$T_{0}$ ($HJD$+2400000)                    & 43392.4858(43)     \\
$P_{orb}$ (days)                           & 1.7011469(5)        \\
$Q$  (days)                                & -6.15(1)$\times$ 10$^{-10}$    \\
$a_{12}sini'$ (AU)                         & 2.99(0.65)               \\
$e'$                                       & 0.31(10)            \\
$\omega'$ (deg)                            & 189(28)               \\
$T'$ (HJD) $+2400000$                      & 34867(1526)        \\
$P_{12}$ (years)                           & 71.1(5.7)             \\
$f(M_{3})$ (M$_\odot$)                     & 0.0052878(158)        \\
$M_{3}$ (M$_\odot$) for $i'$=90$^{\circ}$  & 0.49       \\
$K_{12}$ (km s$^{-1}$)  & 1.3       \\
\hline
\end{tabular}
$$
\end{table}

\begin{figure}[!ht]
\begin{center}
\includegraphics[width=83mm,height=100mm]{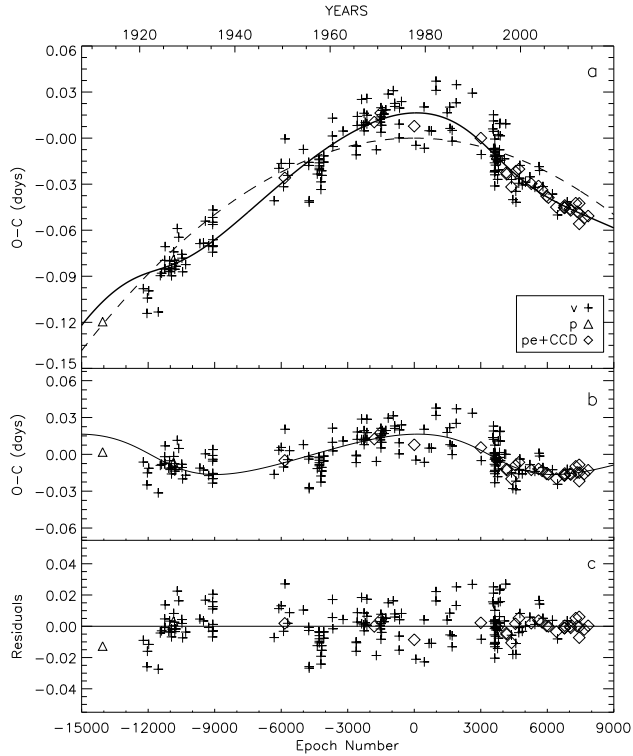}
\caption{(a) Variation of O-C values of W UMi fitted with a
parabolic term (dashed line) and parabolic plus LITE (solid line)
fit. (b) LITE representation of O-C data after removing parabolic
term. (c) Residuals from theoretical representation.} \label{fig4}
\end{center}
\end{figure}

Cyclic O-C changes for W UMi can be also interpreted as possible
magnetic activity on the surface of the cooler companion. \citet{b1}
proposed a model to interpret this type of variation in the periods
of close binaries, which was applied to many active close binaries,
including at least one star with an outer convective envelope. We
used the basic sinusoidal equation to model the cyclic changes of
the O-C values of the target as applied to many active binary stars
\citep[e.g.][]{b46,b56}. Using the method of differential
correction, the parameters of the sinusoidal changes were derived
and are listed in Table 6, together with the Applegate model's
parameters. The rate of the modulation period, resulting from the
magnetic cycle of the cooler, less massive star in the system was
calculated as \emph{ $\Delta$P/P}\,$\approx$\,5.5$\times$10$^{-6}$,
while the average sub-surface magnetic field of the companion of
late spectral type was estimated to be $\approx$2.7 kG. The model
parameters in Table 6 for W UMi were found to be consistent with
those proposed by \citet{b1}.

The value of the quadratic term (\textit{Q}) is indicative of a
decline in the period of the target, with a rate of \emph{dP/dt}=
-1.55 $\times$ 10$^{-7}$ yr$^{-1}$ due to mass loss from the close
binary. Under the assumption of mass transfer from the less massive
component to the more massive one at a rate of 10$^{-8}$ M$_\odot$
yr$^{-1}$ due to the semi-detached status of W UMi, the amount of
mass escaping from the system can be estimated to be \emph{dM/dt}=
-5.8 $\times$ 10$^{-8}$ M$_\odot$ yr$^{-1}$. The calculation method
for the rates of mass loss, together with mass transfer in the
system, can be seen in the study of \citet{b43}.

\begin{table}[!ht]
\caption{Parameters of sinusoidal O-C change and applied Applegate
model for cooler companion in W UMi.} \label{table5}
    $$
        \begin{tabular}{ll}
\noalign{\smallskip} \hline \noalign{\smallskip}
Parameters                      & Value              \\
\noalign{\smallskip} \hline \noalign{\smallskip}
$T_{0}$ ($HJD$+2400000)                    & 43392.4841(54)        \\
$P_{orb}$ (days)                           & 1.7011525(6)         \\
$T_{s}$ (cycle)                            & 3789(30)             \\
$A_{mod}$ (days)                           & 0.016(3)            \\
$P_{mod}$ (days)                           & 25385(75)            \\
$\Delta$\emph{P }(s)                              & 0.582 \\
$\Delta$\emph{P/P}                                & 4.0$\times$10$^{-6}$ \\
$\Delta$\emph{J} ($g\,\,cm^{2}s^{-1}$)                & 8.3$\times$10$^{47}$ \\
$I_{s}$ ($g\,\,cm^{2}$)                        & 8.9$\times$10$^{54}$ \\
$\Delta\Omega/\Omega$                      & 0.0015                \\
$\Delta$\emph{E} (erg)                            & 1.6$\times$10$^{41}$ \\
$\Delta$$L_{rms}$ (erg)                    & 2.2$\times$10$^{32}$ \\
\emph{B} (kG)                                     & 2.7                   \\
$\Delta$m (mag)                            & 0.0006                \\
 \hline
\end{tabular}
$$
\end{table}

\section{Absolute properties and evolutionary status}

The fundamental astrophysical parameters of the semi-detached binary
W UMi were determined with the simultaneous solution of \emph{BVR}
multi-color LCs and the RV data of the companion stars. Although W
UMi is classified as a semi-detached Algol-type binary, a cooler,
less massive star fills its Roche volume as expected, but another
hotter component fills about 85 \% of its Roche lobe according to
the photometric solution. Therefore, the system should be considered
as a near-contact semi-detached binary. The absolute parameters,
which were computed using the parameters listed in Table 3, are
shown in Table 7. The bolometric correction values of the companions
used in the calculation were chosen from \citet{b6} with a solar
temperature (\emph{T$_eff,_\odot$}) of 5777 K and solar bolometric
magnitude (\emph{M$_{bol,_{\odot}}$}) of 4.74 mag.

From the simultaneous solution of the RVs of the companions and
multi-color photometric data, the mass and radius of the hotter
primary star were calculated to be 3.22$\pm$0.08 \emph{M$_{\odot}$}
and 3.63$\pm$0.04 \emph{R$_{\odot}$}, respectively. The designated
mass value corresponds to the B8 spectral type according to the
absolute parameters of main sequence stars by \citet{b7}. The radius
value of the primary component was identified as being fairly large
for the B8 spectral type. The log \emph{g} value of the secondary
star and its other basic parameters indicate that it is a distorted
sub-giant star, as confirmed by photometric results.

\begin{table*}[!ht]
  \caption{Main physical parameters of W UMi.}
  \label{tab:par}
  \begin{center}
           \begin{tabular}[h]{lcc}
  \hline\hline
     Parameter                                 &    Primary    & Secondary  \\
  \hline
     Mass (\emph{M$_{\odot}$})                  & 3.22$\pm$0.08 &1.44$\pm$0.05 \\
     Radius (\emph{R$_{\odot}$})                & 3.63$\pm$0.04 &3.09$\pm$0.03  \\
     Temperature (K)                            & 9310$\pm$90$^{a}$  &5240$\pm$200  \\
     log \emph{L} (\emph{L$_{\odot}$})          & 1.95$\pm$0.03 &0.81$\pm$0.08 \\
     log \emph{g} (cgs)                         & 3.83$\pm$0.02 &3.62$\pm$0.02 \\
     \emph{M$_{bol}$} (mag)                     & -0.13$\pm$0.06 & 2.71$\pm$0.15 \\
     \emph{BC} (mag)                            & -0.24$^{b}$         &    -0.35$^{b}$  \\
     \emph{M$_{v}$} (mag)                       & -0.01$\pm$0.09& 2.92$\pm$0.15           \\
     \emph{E(B-V)} (mag)                        & \multicolumn{2}{c}{0.10}         \\
     Orbital separation (\emph{R$_{\odot}$})    & \multicolumn{2}{c}{10.01$\pm$0.06}          \\
     Photometric distance (pc)                              & \multicolumn{2}{c}{422$\pm$40}   \\
     \emph{Gaia}-DR2 distance (pc)                          & \multicolumn{2}{c}{414$\pm$5}   \\
  \hline
          \end{tabular}\\
      \footnotesize $^{a}$\citet{b24}, $^{b}$\citet{b6}
     \end{center}
    \end{table*}

The photometric distance of W UMi was found to be 422$\pm$40 pc
using the absolute parameters and interstellar reddening given in
Table 7. We estimated the value of the color excess, \emph{E(B-V)}
value to be $\approx0^{m}$.10 from the galactic extinction maps of
\citet{b35}. The calculated distance is compatible with the value,
414$\pm$5 pc, given by the \emph{Gaia}-DR2 database \citep{b10}
within error limits.

We also calculated the components of space velocities (\emph{U,V,W})
based on the proper motions and parallax value from \emph{Gaia}-DR2
database \citep{b10}, and also the velocity of the mass-center of
the system from this study. Differential corrections for the space
velocities were calculated using the formulas given by \citet{b21},
while the corrections for the local standard of rest were according
to the study of \citet{b3}. The corrected space velocity values were
found to be (\emph{U,V,W}) = (-27.9$\pm$0.7, -16.6$\pm$0.9,
-2.2$\pm$1.2) km s$^{-1}$, while the total space velocity of W UMi
was calculated as 32.5$\pm$1.7 km s$^{-1}$. The basic kinematic
characteristics of the system indicate that it may be a member of
the thin disk of our galaxy, according to the criteria of
\citet{b19}.

In order to investigate the evolutionary states of the near-contact
binary W UMi, binary evolutionary models were generated with version
8845 of \texttt{MESA} code (Module for Experiments in Stellar
Astrophysics) using the binary module \citep{b26,b27,b28,b29}.
Models were constructed under two different approaches (conservative
and non-conservative mass transfer) and also the different initial
values of the components' masses (\emph{M$_{1-i}$} and
\emph{M$_{2-i}$}), initial orbital period of the system
(\emph{P$_{orb-i}$}), and also metallicity (Z). Thus, a large number
of models were produced in order to search for compliance with the
observed parameters of the system. During the model generation
process, the initial mass varied between 2.35 and 2.95
\emph{M$_{\odot}$} for the primary component and between 2.35 and
4.45 \emph{M$_{\odot}$} for the secondary component, while we
changed the initial value of the orbital period from 1.3 to 2.3
days. The metallicity value was also altered from Z=0.01 to Z=0.06
for both the conservative and non-conservative models. We also used
parameter $\alpha$, meaning the fractional mass loss from the system
in the vicinity of the secondary component as defined by
\citet{b39}, as a variable during calculations for the
non-conservative models. In addition, magnetic braking effect
\citep{b32} was also used in the model's calculations by considering
the current mass of the cooler component in order to take into
account the angular momentum change of the system during evolution.
Applications of the binary module of \texttt{MESA} code to reveal
the evolutionary state of binary stars can be found in several
studies, which includes information on model calculations and
explanations of the used parameters \citep[e.g.][]{b28,b45,b33}.

\begin{table*}
  \caption{ Evolutionary model parameters derived from \texttt{MESA} for W UMi.}
  \label{tab:par}
  \begin{center}
           \begin{tabular}[h]{lccccc}
  \hline\hline
     Mass transfer    & Z      & \emph{M$_{1-i}$}     & \emph{M$_{2-i}$}     & \emph{P$_{orb-i}$} & Age  \\
     modes            &        & (\emph{M$_{\odot}$}) & (\emph{M$_{\odot}$}) & (days)             & (Myr) \\
  \hline
    Conservative      & 0.014  &  2.36                & 2.35                 &  2.29              & 520 \\
    Conservative      & 0.055  &  2.29                & 2.41                 &  3.10              & 880 \\
    Non-conservative  &0.055   &&&&\\
    $\alpha$=0.1      &        &  2.29          & 2.52                 &  2.98              & 790 \\
    $\alpha$=0.3      &        &  2.37          & 2.73                 &  2.81              & 660 \\
    $\alpha$=0.5      &        &  2.47          & 3.01                 &  2.63              & 530 \\
    $\alpha$=0.7      &        &  2.60          & 3.62                 &  2.27              & 400 \\
    $\alpha$=0.9      &        &  2.95          & 4.45                 &  1.33              & 280 \\
  \hline
          \end{tabular}\\
     \end{center}
    \end{table*}

The possible binary star evolution models, which we can use to
explain the evolutionary status of W UMi, were investigated
comparing the surface gravity (log \emph{g}) and radii of the
companions from the models and analysis of observational data. Model
parameters from \texttt{MESA} are listed in Table 8, which are given
under the approaches of conservative and non-conservative mass
transfer. As seen from the table, we calculated models for solar
metallicity (Z=0.014) and also for metal-rich mode with Z=0.055,
which gives a better match with the parameters of the target derived
from analysis of the observations. In Figure 5, the locations of
both companions of the system in the diagram of log log $L$ - log
\emph{T$_{eff}$} together with the evolutionary models from the
\texttt{MESA} binary module with solar metallicity and also Z=0.055
were produced under the assumptions of conservative and
non-conservative evolution. It can be seen from Figure 5a that the
models produced for solar metallicity do not show good compatibility
with the current properties of the companions of W UMi. Figure 6
indicates the radii variations and also mass-luminosity relations of
the components during non-conservative evolution for Z=0.055.
Evolutionary tracks in Figures 5 and 6 were generated by
\texttt{MESA} for the initial masses of the components in Table 8.

In the non-conservative approach, models were produced for different
values of the fractional mass loss parameter ($\alpha$). The best
match with the current parameters of the system was $\alpha$ = 0.7.
In that case, the system's age was estimated as $\approx$400 Myr,
while conservative models with Z=0.055 give an age of $\approx$880
Myr.

\begin{figure}[!ht]
\begin{center}
\includegraphics[width=83mm, height=120mm]{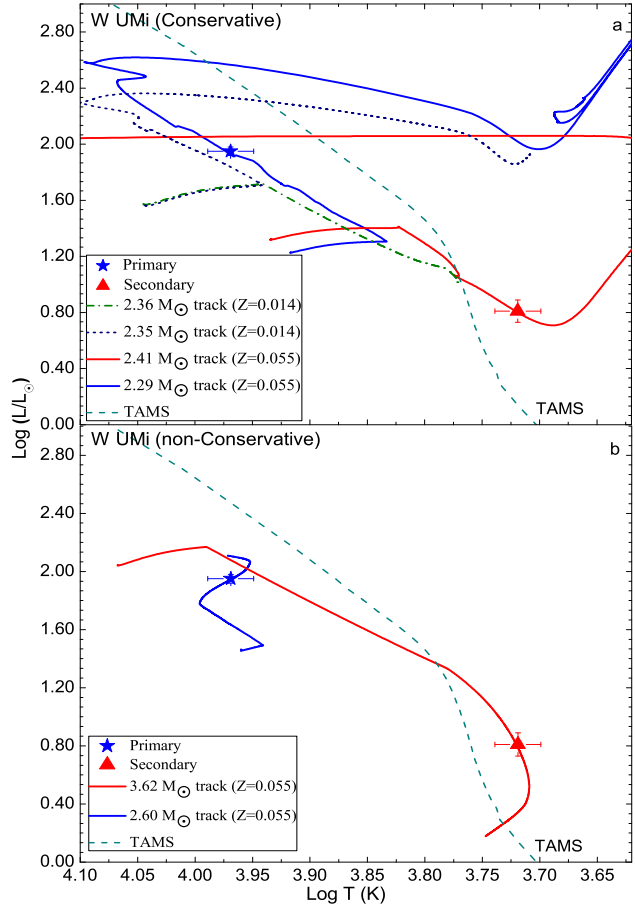}
\caption{Positions of primary and secondary components of W UMi
shown by filled star (primary) and triangle (secondary) in plane of
log $L$ - log \emph{T$_{eff}$}. Dashed lines indicate Terminal Age
Main Sequence (TAMS). Under conservative mass transfer approach,
evolutionary tracks for obtained initial masses according to
metallicities Z=0.014 (dotted-dashed line for primary, dotted line
for secondary) and Z=0.055 (continuous blue line for primary and red
line for secondary) are shown in upper panel (a). In bottom panel
(b), evolutionary tracks calculated under the non-conservative mode
for primary (red line) and secondary (blue line) components are
drawn for Z=0.055. All theoretical curves were calculated by
\texttt{MESA}.} \label{fig5}
\end{center}
\end{figure}

\begin{figure}[!ht]
\begin{center}
\includegraphics[width=83mm, height=120mm]{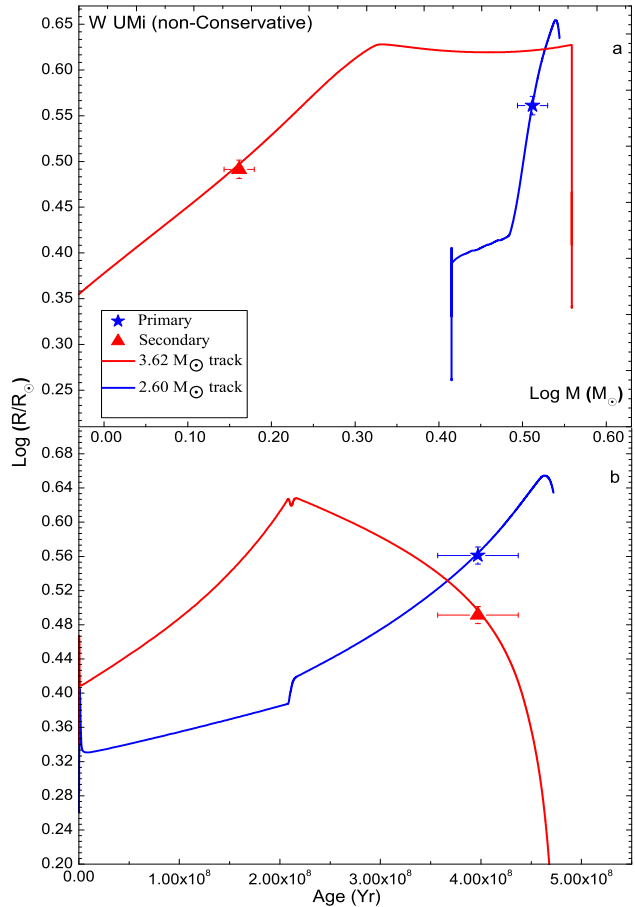}
\caption{Locations of primary (filled star) and secondary (filled
triangle) companions of W UMi in planes of log \emph{R} - log
\emph{M} and log \emph{R} - \emph{Age} in upper (a) and bottom (b)
panels, respectively. Evolutionary calculations for primary
companion with initial mass of 2.60 \emph{M$_{\odot}$} (blue line)
and secondary companion with initial mass of 3.62 \emph{M$_{\odot}$}
(red line) were made for the non-conservative mass transfer mode.
All theoretical curves were calculated by \texttt{MESA}.}
\label{fig6}
\end{center}
\end{figure}

\section{Discussion and Conclusions}

SDB type binaries are valuable for investigation of the evolutionary
phases of close binaries and the processes affecting evolution, such
as mass transfer and loss. On the other hand, it is not easy to
obtain the absolute parameters of these systems due to faint
secondaries in the optical region and also mass transfer and loss
events. In the catalogue by \citet{b13} there are only 38 eclipsing
double-lined SDBs, and so far the number of these systems has not
greatly increased, although many optical LCs of the SDBs have been
gained from several ground-based surveys (e.g. ASAS, SuperWASP) and
satellites (e.g. \emph{Hipparcos}, \emph{Kepler}, \emph{TESS}).

W UMi is a double-lined semi-detached binary and is also close to
the contact phase, as indicated by recent analysis. The target can
therefore be classified as a near-contact, semi-detached binary
based on the description of \citet{b57} because the hotter component
appears to fill approximately 85\% of its inner Roche lobe. It is
worthwhile studying W UMi from this aspect and the reasons are as
follows: i) to derive the main absolute parameters of the companion
stars based on both spectroscopic and photometric analysis, ii) to
investigate its orbital period variation, and iii) to examine the
evolution of the system by producing binary evolution models under
both the conservative and non-conservative mass transfer approach.

The simultaneous analysis of LCs and RV data gives a mass ratio
value (\emph{q}=0.447$\pm$0.005) greater by $\sim$8\% than the value
given by \citet{b24}, which is due to the different measurements of
RV data of the components. Therefore, it can be said that the mass
of the primary component is found to be smaller but its radius is
larger, when we compare them with the values of \citet{b24}. The
errors calculated for the masses and radii of both components are in
the order of 3\% and 1\%, respectively. From the spectral synthesis
analysis, it may be suggested that the system has a higher
metallicity ([Fe/H]$\approx$0.5$\pm$0.2) than the Sun, as seen in
Figure 2, although solar metallicity was assumed by \citet{b24}. The
metal-rich state of the system is supported by the fact that it is a
thin-disk member, as derived from kinematic analysis as well as by
binary evolution models. The surface gravity values of the cooler
companions indicate that it is a sub-giant star. The hotter, primary
component has a lower log \emph{g} value than values of
main-sequence stars with the same spectral type and mass
\citep[e.g.][]{b7}. The mass of the gainer component is found to be
larger than that of main sequence stars at the same temperature,
which is most likely due to the evolutionary state of the target and
the mass loss and transfer process, seen in the generated
evolutionary models.

Mass transfer in classical SDBs takes place from the less massive
component to the more massive star. Therefore, the orbital period of
these systems is expected to increase. In the case of W UMi, the
orbital period was found to be decreasing at a rate of -0.02 s
yr$^{-1}$ when we analyzed minima times reported in the last
century. The decrease in the period may occur due to mass loss from
the system (5.8 $\times$ 10$^{-8}$ M$_\odot$ yr$^{-1}$). In addition
to the secular changes, the O-C values of the target also show a
cyclical change, which can be interpreted as a possible third body
or magnetic activity on the surface of the active component
\citep[e.g.][]{b40,b54}. In this study, LITE analysis indicated that
a tertiary companion around the close pair with a minimum mass of
$\approx$ 0.49 M$_\odot$ and a period of 71 yrs may cause this kind
of cyclic O-C variation. On the other hand, a significant third
light (\emph{l$_{3}$}) could not be determined during the analysis.
In this study, the parameters obtained by applying parabolic and
LITE terms to the O-C values in general are compatible with those
obtained by \citet{b16}, but various values have been found to be
above the error level for some parameters (e.g. mass and orbital
period of possible third body). Magnetic activity of the sub-giant
companion with a temperature of 5240 K may be the cause of such
cyclic variations in the O-C diagram. Therefore, the Applegate model
was also used to explain the cyclic period variations of W UMi and
the calculated model parameters were found to be compatible with
those proposed in \citet{b1}.

In this study, binary evolutionary models were propagated by
\texttt{MESA} code to try to understand the present evolutionary
status and to investigate the initial state of the system. Both
conservative and non-conservative mass transfer approaches were used
for the solar metallicity and also for the system having higher
metallicity than the Sun. Although the conservative models with
Z=0.05 are not far from representing the observational parameters of
the target system, a better fit with the observational data was
achieved in non-conservative models with Z=0.05 values. In some
studies, it was reported that non-conservative models better
represent the observational parameters of classical Algols
\citep[e.g.][]{b36,b51}. The decreasing orbital period of W UMi
determined by O-C analysis, which can also be seen for several
Algols \citep[e.g.][]{b31,b44}, may also be considered as evidence
to support non-conservative evolutionary models.

\begin{figure}[!ht]
\begin{center}
\includegraphics[width=83mm, height=68mm]{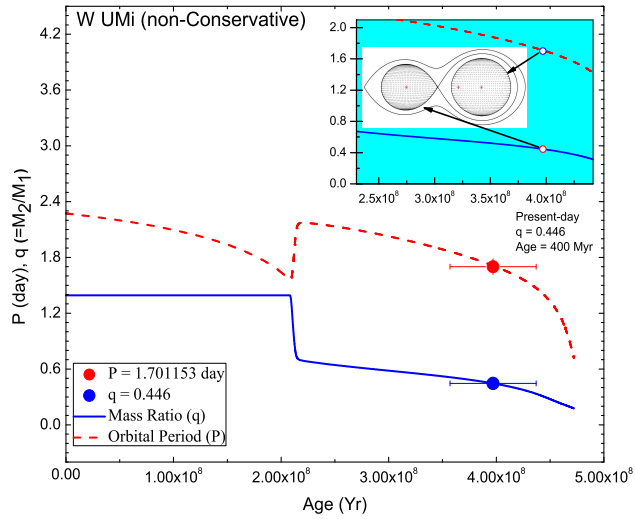}
\caption{Variation of orbital period (P) and mass ratio (q) of W UMi
versus its age until system reaches contact phase. Position of
system based on present P and q values, and also current Roche
geometry (inner box) are indicated. Theoretical lines were
propagated using \texttt{MESA} binary models.} \label{fig7}
\end{center}
\end{figure}

In the most compatible evolutionary models in non-conservative mode,
the age of W UMi could be $\approx$400 Myr with an initial orbital
period of 2.3 days, as seen in Table 8. Important transition points
for the evolution were also examined during the model calculations.
The best non-conservative model matching the observed data suggests
that the first RLOF (Roche-lobe overflow) started when the age of
the target was $\approx$200 Myr and it had an orbital period of
$\approx$ 1.4 days. The mass transfer rate in the system was
calculated to be 1 $\times$ 10$^{-7}$ M$_\odot$ yr$^{-1}$ during the
first RLOF phase. In Figure 7, variation of the orbital period (P)
and mass ratio (q) of the companion stars can be seen during
evolution until the contact phase. In addition, the current P and q
values are indicated together with the present Roche geometry in
Figure 7. Non-conservative binary models show that the near-contact
system W UMi may be a contact binary after about 40-50 Myr. It is
estimated from the models that the values of P and q of the system
will decrease until the contact phase (Fig. 7) and when it reaches
the contact phase, the orbital period and mass ratio will be
$\approx$1.4 days and $\approx$0.3, respectively.

From the results in this study and also the literature, it is
thought that investigation of W UMi type near-contact semi-detached
systems is important to understand the nature of these binaries and
also possible transition from semi-detached status to the contact
phase during the evolution of close interacting binary stars. In the
case of W UMi, more spectroscopic data with high-resolution and S/N
ratios is needed to derive reliably the metal abundance of the
system, to obtain clues for its mass flow and loss processes and to
check the possible magnetic cycle of the cooler component and any
unseen companion around the binary system.

\begin{acknowledgements}
This study was supported by \c{C}anakkale Onsekiz Mart University
the Scientific Research Coordination Unit, Project number:
FBA-2018-2549 and also partly supported by T\"{U}B\.{I}TAK
(Scientific and Technological Research Council of Turkey) under
Grant No. 111T224. The authors wish to thank the all staff of the
Catania Astrophysical Observatory for the allocation of telescope
time and help during observations. This research made use of VIZIER
and SIMBAD databases at CDS, Strasbourg, France. We are grateful to
Dr. Jang-Ho Park for providing us with spectral data of the system.
This work has made use of data from the European Space Agency (ESA)
mission \emph{Gaia} (https://www.cosmos.esa.int/gaia), processed by
the \emph{Gaia} Data Processing and Analysis Consortium (DPAC,
https://www.cosmos.esa.int/web/gaia/dpac/consortium). Funding for
the DPAC has been provided by national institutions, in particular
the institutions participating in the \emph{Gaia} Multilateral
Agreement.
\end{acknowledgements}

\label{lastpage}


\begin{thebibliography}{99}


\bibitem[\protect\citeauthoryear{Applegate}{1992}]{b1} Applegate, J.H. 1992, \apj, 385, 621

\bibitem[\protect\citeauthoryear{Astbury}{1913}]{b2} Astbury, T.H. 1913, AN, 194, 413

\bibitem[\protect\citeauthoryear{Co\c{s}kuno\u{g}lu et al.}{2011}]{b3} Co\c{s}kuno\v{g}lu,
B., Ak, S., Bilir, S. et al., 2011, \mnras, 412, 1237

\bibitem[\protect\citeauthoryear{Devinney et al.}{1970}]{b4} Devinney, E. J., Hall, D.S. and Ward, D.H.,
1970, \pasp, 82, 10

\bibitem[\protect\citeauthoryear{Djurasevic et al.}{2003}]{b5}
Djurasevic, G., Rovithis-Livaniou, H., Rovithis, P. et al., 2003,
\aap, 402, 667

\bibitem[\protect\citeauthoryear{Drilling \& Landolt}{2000}]{b6} Drilling John S. \& Landolt A. U.,
2000, Allen's Astrophysical Quantities, 4th ed., Edited by Arthur N.
Cox. ISBN: 0-387-98746-0. Publisher: New York: AIP Press, Springer

\bibitem[\protect\citeauthoryear{Eker et al.}{2018}]{b7} Eker, Z., Bak{\i}\c{s},
V., Bilir, S., Soydugan, F., Steer, I., Soydugan, E., Bak{\i}\c{s},
H., Ali\c{c}avu\c{s}, F., Aslan, G. and Alpsoy, M., 2018, \mnras,
479, 5491

\bibitem[\protect\citeauthoryear{Frasca et al.}{2000}]{b8} Frasca, A.,
Marino, G., Catalano, S. and Marilli, E., 2000, \aap, 358, 1007

\bibitem[\protect\citeauthoryear{Gaia Collaboration et al.}{2016}]{b9}
Gaia Collaboration, Brown, A. G. A., Vallenari, A. et al., 2016,
\aap, 595, 2

\bibitem[\protect\citeauthoryear{Gaia Collaboration et al.}{2018}]{b10}
Gaia Collaboration, Brown, A.G.A., Vallenari, A., et al., 2018,
\aap, 616, A1

\bibitem[\protect\citeauthoryear{Joy \& Dustheimer}{1935}]{b11} Joy, A.H. and Dustheimer, O.L.,
1935, \apj, 81, 479

\bibitem[\protect\citeauthoryear{Irwin}{1959}]{b12} Irwin, J.B., 1959, \aj, 64, 149

\bibitem[\protect\citeauthoryear{\.{I}bano\u{g}lu et al.} {2006}]{b13} \.{I}bano\u{g}lu C., Soydugan F.,
Soydugan, E. \& Dervi\c{s}o\u{g}lu A., 2006, \mnras, 373, 435

\bibitem[\protect\citeauthoryear{Kaitchuck \& Honeycutt}{1982}]{b14} Kaitchuck,
R.H., Honeycutt, R.K., \pasp, 94, 532

\bibitem[\protect\citeauthoryear{Kreiner at al.}{2001}]{b15} Kreiner, J.M., Kim, C.-H, Nha, I.-S.,
2001, An Atlas of O-C Diagrams of Eclipsing Binary Stars, Parts 1-6,
Cracow: Pedagogical University Press

\bibitem[\protect\citeauthoryear{Kreiner et al.}{2008}]{b16}
Kreiner, J.M., Pribulla, T., Tremko, J., Stachowski, G.S. and
Zakrzewski B., 2008, \mnras, 383, 1506

\bibitem[\protect\citeauthoryear{Kreiner}{2004}]{b17} Kreiner, J.M., 2004, \actaa, 54, 207

\bibitem[\protect\citeauthoryear{Kurucz}{1993}]{b18} Kurucz, R.L., 1993, Peculiar versus Normal
Phenomena in A-type and Related Stars. International Astronomical
Union, Colloquium No. 138, Italy, Editors, M.M. Dworetsky, F.
Castelli, R. Faraggiana; Publisher, Astronomical Society of the
Pacific, San Francisco, CA, Vol. 44, p87

\bibitem[\protect\citeauthoryear{Leggett}{1992}] {b19} Leggett,
S.K., 1992, \apjs, 82, 351

\bibitem[\protect\citeauthoryear{Lucy}{1967}]{b20} Lucy, L.B., 1967, \zap, 65, 89

\bibitem[\protect\citeauthoryear{Mihalas \& Binney}{1981}]{b21} Mihalas, D.
and Binney, J., 1981, Galactic astronomy: Structure and kinematics,
2nd edition, San Francisco, CA, W. H. Freeman and Co., p.608

\bibitem[\protect\citeauthoryear{Mardirossian et al.}{1980}]{b22} Mardirossian, F., Mezzetti, M.,
Predolin, F. and Giuricin, G., 1980, A\&AS, 40, 57

\bibitem[\protect\citeauthoryear{Nakamura et al.}{1998}]{b23} Nakamura,
Y., Asada, K. and Sato, R., 1998, IBVS, 4647, 1


\bibitem[\protect\citeauthoryear{Park et al.}{2018}]{b24} Park, J.H.,
Hong, K., Koo, J.-R., Lee, J.W., Kim, C.-H., 2018, \aj, 155, 133

\bibitem[\protect\citeauthoryear{Payne-Gaposchkin}{1952}]{b25}
Payne-Gaposchkin, C., 1952, Ann. of Harvard College Obser., 118, 217

\bibitem[\protect\citeauthoryear{Paxton et al.}{2011}]{b26}
Paxton, B., Bildsten L., Dotter, A., Herwig, F., Lesaffre, P.,
Timmes, F., 2011, \apjs, 192, 3

\bibitem[\protect\citeauthoryear{Paxton et al.}{2013}]{b27}
Paxton B. et al., 2013, \apjs, 208, 4

\bibitem[\protect\citeauthoryear{Paxton et al.}{2015}]{b28}
Paxton B. et al., 2015, \apjs, 220, 15

\bibitem[\protect\citeauthoryear{Paxton et al.}{2018}]{b29}
Paxton B. et al., 2018, \apjs, 224, 34

\bibitem[\protect\citeauthoryear{Piskunov et al.}{1995}]{b30}
Piskunov, N. E., Kupka, F., Ryabchikova, T. A., Weiss, W. W.,
Jeffery, C. S., 1995, A\&AS, 112, 525

\bibitem[\protect\citeauthoryear{Qian}{2002}]{b31} Qian, S., 2002,
\pasp, 114, 650

\bibitem[\protect\citeauthoryear{Rappaport et al.}{1983}]{b32} Rappaport,
S., Verbunt, F. and Joss, P.C., 1983, \apj, 275, 713

\bibitem[\protect\citeauthoryear{Rosales et al.}{2019}]{b33} Rosales,
J.A., Mennickent, R.E., Schleicher, D.R.G., Senhadji, A.A., 2019,
\mnras, 483, 862

\bibitem[\protect\citeauthoryear{Rucinski}{1969}]{b34}Rucinski, S. M., 1969, \actaa, 19, 245

\bibitem[\protect\citeauthoryear{Schlegel et al.} {1998}] {b35} Schlegel, D.J., Finkbeiner, D.P.,
Davis, M., 1998, \apj, 500, 525


\bibitem[\protect\citeauthoryear{Sarna}{1993}]{b36} Sarna, M.J.,
1993, \mnras, 262, 534

\bibitem[\protect\citeauthoryear{Shade}{1945}]{b37} Sahade, J., 1945, \apj, 102, 470

\bibitem[\protect\citeauthoryear{Shaw}{1994}]{b38} Shaw, J.S., 1994, \memsai, 65, 95


\bibitem[\protect\citeauthoryear{Soberman et al.}{1997}]{b39} Soberman,
G.E., Phinney, E.S. and van den Heuvel, E.P.J., 1997, \aap, 327, 620

\bibitem[\protect\citeauthoryear{Soydugan et al.}{2003}]{b40}Soydugan, F., Demircan, O., Soydugan, E.,
\.{I}banoglu, C., 2003, \aj, 126, 393

\bibitem[\protect\citeauthoryear{Soydugan et al.}{2006}]{b41}Soydugan, F., Soydugan, E.,
\.{I}banoglu, C., Demircan, O., 2006, AN, 327, 705

\bibitem[\protect\citeauthoryear{Soydugan et al.}{2007}]{b42}Soydugan, F., Frasca, A., Soydugan,
E., Catalano, S., Demircan, O., \.{I}bano\v{g}lu, C., 2007, \mnras,
379, 1533

\bibitem[\protect\citeauthoryear{Soydugan et al.}{2011a}]{b43}Soydugan,
F., Erdem, A., Dogru, S.S., Ali\c{c}avu\c{s}, F., Soydugan, E.,
\c{C}i\c{c}ek, C., Demircan, O., 2011a, \na, 16, 72

\bibitem[\protect\citeauthoryear{Soydugan et al.}{2011b}]{b44}Soydugan,
E., Soydugan, F., \c{S}eny\"{u}z, T., Puskullu, C., Demircan, O.,
2011b, \na, 16, 72

\bibitem[\protect\citeauthoryear{Streamer et al.}{2018}]{b45} Streamer, M.,
Ireland, M.J., Murphy, S.J., Bento, J., 2018, \mnras, 480, 1372

\bibitem[\protect\citeauthoryear{Tian et al.}{2009}]{b46} Tian, Y.P.,
Xiang, F.Y. and Tao, X., 2009, \apss, 319, 119

\bibitem[\protect\citeauthoryear{von Ziepel}{1924}]{b47} von Ziepel, H., 1924, \mnras, 84, 665

\bibitem[\protect\citeauthoryear{Wilson $\&$ Devinney}{1971}]{b48} Wilson, R.E., Devinney, R.J.,
1971, \apj, 166, 605

\bibitem[\protect\citeauthoryear{Valenti $\&$ Piskunov}{1996}]{b49} Valenti,
J.A. and Piskunov, N., 1996, A\&AS, 118, 595

\bibitem[\protect\citeauthoryear{van Hamme}{1993}]{b50}
van Hamme W., 1993, \aj, 106, 2096

\bibitem[\protect\citeauthoryear{van Rensbergen et al.}{2011}]{b51} van
Rensbergen, W., Mennekens, N., de Greve, J.-P., Jansen, K., de
Loore, B., 2011, \aap, 582, 16

\bibitem[\protect\citeauthoryear{Yang}{2013}]{b52} Yang, Y.-G.,
2013, \na, 25, 109

\bibitem[\protect\citeauthoryear{Zasche et al.}{2008}]{b53} Zasche, P.,
Liakos, A., Wolf, M. and Niarchos, P., 2008, \na, 13, 405

\bibitem[\protect\citeauthoryear{Zasche et al.}{2010}]{b54} Zasche, P.,
Uhl\'{a}\u{r}, R., Svoboda, P., 2010, \apss, 326, 119

\bibitem[\protect\citeauthoryear{Zasche et al.}{2014}]{b55} Zasche, P.,
Wolf, M., Uhla\u{r}, R., Ku\u{c}\'{a}kov\'{a}, H., 2014, \aj, 147,
130

\bibitem[\protect\citeauthoryear{Zhang et al.}{2019}]{b56} Zhang, B., Qian,
S.-B., Zhi, Q.-J. et al., 2019, \pasp, 131, 034201

\bibitem[\protect\citeauthoryear{Zhu \& Qian}{2009}]{b57} Zhu, L., Qian, S., 2009, The Eighth Pacific
Rim Conference on Stellar Astrophysics, edited by B. Soonthornthum,
S. Komonjinda, K.S. Cheng, and K.C. Leung San Francisco:
Astronomical Society of the Pacific, Vol. 404, p.189



\end{thebibliography}
\end{document}